\documentclass[prb,aps,showpacs,preprintnumbers,amsmath,amssymb]{revtex4}
\usepackage{amsmath,amssymb,array}

\begin{document}

\def\brho{{\mbox{\boldmath $\rho $}}}
\def\bomega{{\mbox{\boldmath $\Omega $}}}
\def\bomicron{{\mbox{\boldmath $\omega $}}}
\def\ul#1#2{\textstyle{\frac#1#2}}

\title{Density waves theory of the capsid structure of small icosahedral viruses} 

\author{V.L. Lorman$^\ddagger$ and S.B. Rochal$^{\dagger,\ddagger}$}

\noindent

\affiliation{$^\ddagger$Laboratoire de Physique  Theorique  et Astroparticules, CNRS - Universite
Montpellier 2, Place Eugene Bataillon, 34095 Montpellier, France \\
$^\dagger$Physical Faculty, Rostov State University, 5 Zorge Str.,
344090 Rostov-on-Don, Russia}


\noindent

\begin{abstract}
\small
We apply Landau theory of crystallization to explain and to
classify the capsid structures of small viruses with spherical
topology and icosahedral symmetry. We develop an explicit method
which predicts the positions of centers of mass for the proteins
constituting viral capsid shell. Corresponding density
distribution function which generates the positions has universal
form without any fitting parameter.  The theory describes in a
uniform way both the structures satisfying the
well-known Caspar and Klug geometrical model for capsid
construction and those violating it. The quasiequivalence of
protein environments in viral capsid and peculiarities of the
assembly thermodynamics are also discussed.
\end{abstract}
\maketitle

Viruses represent rather simple biologocal systems which can be
studied by different chemical and physical methods. Their
organization and functionning show a number of universal features.
The viral protein shell (capsid) encloses the genetic material (either
desoxyribonucleic acid (DNA) or ribonucleic acid (RNA) [1])
responsible for the infective properties of the viruses. The capsid
serves both to preserve and to transmit the genetic material to an
appropriate host cell. Soon after the transmission the host cell
starts the reproduction of the viral DNA (or RNA) and capsid
proteins. From shell proteins and replicated genomes, new
identical copies of the viruses spontaneously assemble.
Though the final infective virus structure formation involves
biologically specific events, some steps of the self-assembly demonstrate
properties typical for ordering in passive physical systems.
The host cell is not necessary for the viral capsid formation.
The self-assembly does not need active local energy consumption
like ATP hydrolisis and the process can be reversible [2,3].
Moreover, in many cases the viral shells assembling does
not even need genomes and proceeds in vitro in purified protein solutions [1].

The problem of the capsid structure formation attracts the
attention of physicists since fifty years. In their pioneer work
Crick and Watson [4] stated that spherical viruses should have the
symmetry (but not necessarily the structure) of one of regular
polyhedra with the faces formed by identical perfect polygons.
Later in 1962, Caspar and Klug (CK) argued that spherical capsids
adopt icosahedral point symmetry [5]. They have seen the physical
reason why the Nature prefers this type of symmetry in the fact
that the icosahedron has the largest volume-to-surface ratio among
the regular polyhedra. Besides, CK obtained four prominent results
[5]: i) The capsid symmetry is lower than that of the regular
icosahedron since the proteins are asymmetric. Identical asymmetric
building blocks can compose the structures with rotational symmetry
elements only, excluding inversion and mirror planes. ii) The
asymmetric proteins can be located only in regular (trivial)
60-fold positions of the rotational icosahedral point group,
therefore the total number of proteins in a capsid is always equal
to 60N, where N is a positive integer number. iii) CK concluded,
for the first time, that 'the self-assembly is a process akin to
crystallisation and is governed by the laws of statistical
mechanics'. iv) They proposed a geometrical model for the viral
capsid construction based on the properties of the almost regular
mapping of the 2D hexagonal structure on the icosahedron surface.
Specific properties of the model impose the selection rules for the
value of N (and, consequently, for the total number of proteins in
the shell). Only the values which satisfy the relation $N= h^2+
k^2+hk$, where h and k are non-negative integers are allowed by the
CK selection rules. All four points and their direct consequences
resulted in the principle formulated by CK and
put in the basis of modern virology. Though a big number of virus
capsid structures are in a good agreement with all the points of
the CK scheme, there is a growing number of experimentally resolved
structures which do not satisfy the CK selection rules nor their
predictions about local proteins arrangment [6]. These facts show that point iv) of the
principle is not universal and needs to be generalized.

In recent years the investigation of capsid structures has undergone a real
burst due to the progress of the X-ray and cryoelectron microscopy
techniques and micromechanical experiments [7].
From the theoretical point of view the main
effort was done in two directions (see [8]). On the one hand, the
mean-field studies of simple model systems were performed
in order to approach the thermodynamics of the self-assembly process.
On the other hand, the mechanical properties of capsids and their
relation to the capsid shape were investigated. Along the first line,
the free energy of the viral structure has been approximated by that
of a model system consisting of two types of disks located on the
spherical surface [9]. The proposed pair potential of the disk
interaction favors the icosahedral symmetry of the disk packing [9]
provided an optimisation of several model parameters. Along the second line,
the possible buckling instability of the spherical capsid structure
was studied in the frame of the nonlinear physics of thin elastic
shells [10]. The results of this study explain why
the relatively small viruses are always spherical while the larger
ones have a more angular or faceted shape. In addition, for large
viruses the use of continuum elasticity approximation can be
justified. This makes the predictions of the mechanical properties [11] of
viral capsids and their large-scale shape details [12] more universal.
Nevertheless, the results obtained depend crucially on the
model assumptions concerning the explicit form of interaction
between proteins or groups of proteins (capsomers). Let us also
note that all recent theoretical works on the capsid structure do not
take into account the asymmetry of capsid proteins nor the restrictions
on the capsid symmetry formulated in points i) and ii) of the CK
principle. By contrast, the nonuniversal CK selection rules (point iv))
are taken as an ingredient in all models.

In the present work we propose to apply the Landau theory of
crystallization to the problem of small capsid formation. Resulting
approach to the icosahedral virus structure accounts explicitly for
the protein symmetry and satisfies points i)-iii) of the CK principle but it is free of
nonuniversal CK selection rules. It allows us to describe in a
uniform way all experimentally observed small spherical viruses
including those which can not be obtained using the CK geometrical
model (e.g. L-A virus, Dengue virus, West Nile virus, Murine
Polyoma virus, etc.)

Both the experimental data and the theoretical consideration [10]
show that the shape of small viruses with the icosahedral symmetry
is close to the spherical one. This fact gives the possibility to
consider the crystallization on a spherical surface and to avoid
the problems arising in the CK geometrical construction during the
mapping of planar hexagonal structures upon the icosahedron
surface. Like in the case of usual 3D crystal solidification [13] Landau
theory of the assembly process gives simple and clear predictions
in the vicinity of crystallization point. In this region the
probability density $\rho$ of protein distribution in the capsid
structure is presented as:
\begin{equation}
\rho= \rho_0+\Delta \rho,
\end{equation}
where $\rho_0$ is an isotropic density in the solution and $\Delta
\rho$ corresponds to the density deviation induced by the ordering.
The symmetry breaking during the crystallisation is associated with
one critical order parameter which spans an irreducible
representation of the symmetry group of the disordered state. In
addition, in the vicinity of crystallization point, the structure
of the ordered state (defined by $\Delta \rho$) is determined by
the critical order parameter only, the contribution of non-critical
degrees of freedom being negligible in this region. For the
crystallization process the order parameter represents a critical
system of density waves (CSDW) with the wave vectors of the
\textit{same length} and the transition free energy is an invariant
function of the CSDW amplitudes [13]. The symmetry of crystals
which condense from the isotropic state coincides exactly with that
of the corresponding CSDW. For crystals of metals (and especially
for the crystals of elements) the atomic positions in the vicinity
of crystallization point can be then associated with the positions
of maxima of the CSDW.

The same
principles are applied here to the assembly process on a sphere. The
critical part $\Delta \rho_l$ of the density is determined by a
CSDW with the \textit{same wave number} $l$. The spherical harmonics
$Y_{lm}$ constituting CSDW on a sphere span one irreducible
representation (IR) of the SO(3) symmetry group of the disordered
state, thus  $\Delta \rho_l$ is given by:

\begin{equation}
\Delta \rho_l(\theta, \phi)= \sum_ {m=-l}^{m=l} A_{lm} Y_{lm}(\theta, \phi),
\end{equation}
where $l$ is the IR number, $A_{lm}$ are the amplitudes of the spherical
harmonics $Y_{lm}$ and $\theta$ and $\phi$ are the conventional angular
variables of the spherical coordinate system.

According to points i) and ii) of the CK principle the ordered
distribution of proteins in the viral capsid has the symmetry group
$I$ of the icosahedron rotations which does not contain spatial
inversion nor mirror planes. This restriction is of major
importance in the proposed theory. It selects the parity of the
'active' IR's of the SO(3) symmetry group which induce the assembly
of icosahedral shells of asymmetric proteins. Thus the spherical
harmonics $Y_{lm}$ with \textit{even} $l$ numbers cannot form critical
density (2) for viral capsids. The restriction affects also the free
energy expansion of the assembly process taken in a standard for
the crystallization theory form [13] $F=F_0+F_2+F_3+F_4+...$ and
containing invariant terms
\begin{eqnarray}
F_2&=&A(T,c)\sum_{m=-l}^{m=l}A_{l,m}A_{l,-m},\nonumber \\
F_3&=&B(T,c)\sum_{m_l,m_2,m_3}a_{m_l,m_2,m_3}
A_{l,m_1}A_{l,m_2}A_{l,m_3}\delta(m_l+m_2+m_3)\equiv 0,\\
F_4&=&\sum_{k}C_k(T,c)\sum_{m_l,m_2,m_3,m_4}a_{m_l,m_2,m_3,m_4}^k
A_{l,m_1}A_{l,m_2}A_{l,m_3}A_{l,m_4}\delta(m_l+m_2+m_3+m_4),
\nonumber
\end{eqnarray}
where $a_i$ are weight coefficients of the SO(3)
group (e.g.  Clebsch-Gordan coefficients for the third order term $F_3$),
$\delta(0)=1$, $\delta(i\neq 0)=0$, $A(T,c)$, $B(T,c)$, and $C^k(T,c)$ are temperature- and
composition-dependent coefficients of the Landau theory. For any
\textit{odd} wave number $l$ the third-order  term $F_3$
is identically zero. This fact makes the thermodynamics of
asymmetric proteins assembly quite different with respect to the
thermodynamics of 3D icosahedral atomic  clusters formation [14]
in spite of several common points in formal description.

Next restriction on the choice of order parameters of the capsid
formation comes from the fact that $\Delta \rho_l$ function with $I$
symmetry can be constructed not for all but for particular odd $l$ numbers only.
The analysis based upon the theory of invariants shows that
any critical order parameter which drives the icosahedral assembly
of asymmetric proteins has the wave number $l$ satisfying the
relation:
\begin{equation}
l=15+6i+10j,
\end{equation}
where $i$ and $j$ are positive integers or zero. Eq. (4) defines
the list of $l$ numbers for which the restriction of an IR of the
SO(3) group on the icosahedral group $I$ contains at least one
totally symmetric representation. The sequence of the permitted
values of the wave number $l$ is given by: $l=(15, 21, 25, 27, 31,
33, 35...)$. As we  show below this sequence determines
possible capsid shell structures for small icosahedral viruses.
Selection rule (4) gives the possibility to obtain the explicit
form of critical density (2). Then the protein centers are associated with the
positions of maxima of $\Delta\rho_l$ function (2). Thus the
density wave approach replaces nonuniversal geometrical model iv)
of the CK principle.

The explicit form of the critical density function $\Delta
\rho_l(\theta, \phi)$ is given by the basis
functions $f_l^i (\theta, \phi)$ ($i=1,2...n_t$) of all $n_t$
totally symmetric representations of the icosahedral group $I$
in the restriction of the 'active' IR of the SO(3). The CSDW
is a linear combination of these functions invariant with respect to the $I$ group:
\begin{equation}
\Delta \rho_l (\theta, \phi)= \sum_ {i=1}^{n_t} B_i f_l^i (\theta,
\phi),
\end{equation}
where $B_i$ are arbitrary coefficients.

 Their number
$n_t$ is equal to the number of integer non-negative solutions
$(i,j)$ of Eq. (4) for a fixed permitted value of $l$. Another way to calculate $n_t$ is to use the
well-known relations of characters [15]:
\begin{equation}
        n_t=1/|G|\sum_G\xi(\hat{g})
\end{equation}
where the sum runs over the elements $\hat{g}$ of the $I$ group,
$|G|=60$ is the $I$ group order, and $\xi(\hat{g})$ is the
character of the SO(3) group element  which reads as [15]:
$$
\xi(l,\alpha)=\frac{sin((l+1/2)\alpha)}{sin (\alpha/2)},
$$
where $l$ is the IR number and the angle $\alpha$ is determined by the element $\hat{g}$.
Then the explicit form of (6) becomes:
\begin{equation}
n_t(l)=\frac{1}{60}(2l+1 +15\xi(l,\pi) +20\xi(l,2\pi/3) +12\xi(l,2\pi/5)
+12\xi(l,4\pi/5)).
\end{equation}

For small icosahedral capsids the practical construction of the
protein density distribution is simplified because the CSDW (5)
contains only one function $f_l(\theta,\phi)$. Indeed, according to
Eq. (4) and/or Eq. (7) $n_t=1$ for all $l\leq 43$. In this simplest case
$\Delta\rho_l(\theta,\phi)=B f_l(\theta, \phi)$, where $B$ is a
single arbitrary coefficient. The positions of maxima of the
density function do not depend on the value of $B$. They are
generated by a single universal function $f_l(\theta, \phi)$ which
has no any fitting parameter. In the following consideration the
functions $f_l(\theta, \phi)$ possessing this properties are called
irreducible icosahedral density functions and the structures
generated by $f_l(\theta, \phi)$ are mentioned as irreducible
icosahedral structures. The explicit form of the irreducible
density function $f_l(\theta, \phi)$ for a given value of $l$ is
obtained by averaging of $Y_{lm}(\theta, \phi)$ harmonics over the
$I$ symmetry group [16].
\begin{equation}
f_l (\theta, \phi) =\frac{1}{60}\sum_{G} Y_{l,m}(\hat{g}(\theta,
\phi)).
\end{equation}
For any fixed value of $m$, procedure (8) gives either the same
function $f_l (\theta, \phi)$ we are looking for, or zero.
Functions which differ by a constant complex multiplier are
considered the same.

Fig. 1 resumes the irreducible density functions $f_l(\theta,
\phi)$ permitted by selection rule (4) for the five smallest
icosahedral capsids (Fig. 1(a-e));  the function $f_{37}(\theta,
\phi)$ (Fig. 1(f)) is added as an example illustrating protein density
distribution with higher $l$. The value of $f_l(\theta,\phi)$ is
represented using false color image: variation of colors from red
to violet corresponds to the function growth. Note that all
$f_l(\theta,\phi)$ functions are anti-symmetric : they change their
sign under the inversion of all coordinates or under the action of
mirror planes of a regular icosahedron. Thus, for the sake of
clariry, we present the positive part $f_l(\theta,\phi)>0$ only.
The number of maxima of the density functions is equal to $60N$,
where $N$ is the number of different regular 60-fold positions of
the $I$ group. In the viral capsid $N$ corresponds to the number of
different positions occupied by the proteins. Let us stress that in
a sharp contrast with the CK geometrical model the crystallization
theory predicts the existence of capsids with all positive integer
values of $N$ and not only for $N=h^2+hk+k^2$. Functions
$f_l(\theta,\phi)$ generate in a uniform way protein distributions
which can be obtained by the CK mapping of the hexagonal lattice on
an icosahedron and those which can not. On the one hand, the
distributions in Fig. 1(a) ($l=15$, $N=1$), Fig. 1(d) ($l=27$,
$N=3$), and Fig. 1(e) ($l=31$, $N=4$) give classical CK structures.
The positions of protein centers in a big number of viral capsids
are described by these structures. Fig. 2 shows the correspondence
between the maxima of $f_{15}$, $f_{27}$, and $f_{31}$ and the
protein arrangement in Satellite Tobacco Mosaic virus (Fig. 2(a)),
Cowpea Chlorotic Mottle virus (Fig. 2(b)) and Sindbis virus (Fig.
2(c)), respectively. On the other hand, the distributions in Fig. 1(b) ($l=21$,
$N=2$) and in Fig. 1(f) ($l=37$, $N=6$) do not satisfy the CK
selection rules for $N$ number. The distribution in Fig. 1(c)
($l=25$, $N=3$) shows no hexagonal arrangements of protein
positions and can not be obtained by the CK geometrical model,
though the number of protein positions $N$ satisfies the CK
selection rules. In addition, the comparison of distributions in
Fig. 1(c) and Fig. 1(d) illustrates another striking result of the
crystallization theory : there exist qualitatively different capsid
structures (induced by $f_l(\theta,\phi)$ functions with different
$l$) but with the same number $N$ of protein positions.

The X-ray and cryomicroscopy data show the existence of a whole
series of viral capsids which violate the CK geometrical model but
correspond to the distributions generated by density functions
$f_l(\theta,\phi)$. Fig 3(a) illustrates the correspondence between
the positions of maxima of $f_{21}(\theta,\phi)$ and the structure
[17] of L-A virus with $N=2$; Fig 3(b) relates the maxima of
$f_{25}(\theta,\phi)$ to the structure [17] of Dengue virus with
$N=3$ and in Fig 3(c) the maxima of $f_{37}(\theta,\phi)$ are
compared with the protein distribution [17] of Murine Polyoma virus
with $N=6$.

As an additional comment note, that in our opinion (contrary to the
opinion of Ref. [18]) the above structures violating the CK
geometrical model do not violate the CK idea about the
quasiequivalence of proteins in the viral capsid. CK stated [5]
that since the proteins are identical their environments in the
viral structure should be similar. Initially, this idea was used by
CK to justify their geometrical model. The hexagonal planar
crystalline structure proposed by CK to be the first step of the model
contains six proteins per unit cell. All
proteins of the structure are symmetry-equivalent since they belong to the same
regular orbit of the corresponding planar symmetry group. On
the second step of the CK model, after the mapping of the
planar structure on the icosahedron surface, the same
proteins belong to different 60-fold orbits of the I-group.
In any 3D icosahedral capsid structure the
proteins which belong to different positions can not be
symmetry-equivalent. Nevertheless, the CK geometrical construction
ensures approximate structural equivalence of proteins from
different orbits. This "quasiequivalence" means that the local
order around any protein (the distances between proteins, the
number of nearest neighbors) is more or less the same. The latter
property is intrinsic not only to the CK structures but also for
the capsid structures violating the CK geometrical model including those shown
in Fig 3. Indeed, each position (location of maximum) in all these
structures have five or six nearest neighbors and the distances to
these neighbors are approximately equal. In other words if the
asymmetrical identical building blocks can be slightly deformed (
it is also assumed in the original CK theory) then there is no
problem to put them together in the structure in slightly different
local environments.

Let us finally briefly discuss particular features of the assembly
thermodynamics. Due to the absence of the cubic term in free energy
(3) the icosahedral capsid assembly can be second order phase
transition. Thermodynamic processes of this type have two
advantages for the assembly optimization: they need no latent heat
to be involved in; and they take place without nucleation process.
The latter feature is confirmed experimentally for a number of
small viruses [19] : at equilibrium, either intact virus shell or
free proteins are dominant species while assembly intermediates
(capsid germs) are found in trace concentration.

We would like to stress that irreducible icosahedral density
function $f_l(\theta,\phi)$ contains much more physical
information than simple positions of proteins centers. The full
density distribution generated by $f_l(\theta,\phi)$ is very
useful for understanding of biologically important properties like
virus infectivity. Recent advances in virology have shown that
that infectivity promoted by interaction of cell receptors with
virus surface depends not only on bio-specific binding properties but
also on the capsid proteins distribution. Along this line, the
relation can be established between the minima of $f_l(\theta,\phi)$
and binding sites on the capsid surface. One-to-one correspondence
of the deepest minima of $f_{25}$ (Fig. 1(c)) and the binding
sites for the carbohydrate recognition domains of the dendritic
cell receptors on the Dengue virus surface [Fig 2 in Ref. 20] can
be taken as an illustration of the relation.

\begin{center}
{\Large Figure captions}
\end{center}
\bigskip

Fig 1. (a)-(e) : The first five irreducible icosahedral density functions
with the wave numbers $l=$15, 21, 25, 27, and 31, respectively. Corresponding numbers of
different 60-fold positions of density maxima are $N=$1, 2, 3, 3, and
4. (f): Function with $l=$37 and $N=$6.

Fig 2. Comparison of the positions of proteins centers predicted by
our model (left panel) with the experimental viral structures [17]
(right panel) for the capsids satisfying selection rules of the CK
geometrical model. Capsids of Satellite Tobacco Mosaic virus (a),
Cowpea Chlorotic Mottle Virus (b), and Sindbis virus (c) are presented. The
corresponding density functions for $l=15$,
$l=27$, and $l=31$, respectively, are shown in Fig. 1(a), (d), and (e).

Fig 3. Comparison of the positions of proteins centers  predicted by our model
(left panel) with the experimental viral structures [17] (right panel)
for the capsids which can not be explained by the CK geometrical
model. Capsids of L-A  virus (a), Dengue
virus (b), Murine Polyoma  virus (c) are presented. The corresponding
density functions for $l=21$, $l=25$, and
$l=37$, respectively are shown in Fig. 1(b), (c), and (f).

\end{document}